\documentclass[%preprint,
twocolumn,
aps,
amsfonts,
amssymb,
nofootinbib,
byrevtex
,tightenlines,
preprintnumbers
]{revtex4}

\usepackage{amscd}
\usepackage{amsmath}
\usepackage{bm}
\usepackage[dvips]{color}                 %%% for colours in the graphs
\usepackage{graphicx}
\usepackage{eucal}
\usepackage{pslatex}

\def\){\right)} 
\def\({\left(} 
\def\]{\right]} 
\def\[{\left[}

\def\Journal#1#2#3#4{{#1} {\bf #2}, #3 (#4)}

\def\PRL{\em Phys. Rev. Lett.}

\def\PRC{{\em Phys. Rev.} C}
\def\PRB{{\em Phys. Rev.} B}
\def\PRA{{\em Phys. Rev.} A}
\def\PR{{\em Phys. Rev.} }

\newcommand{\mcal}[1]{{\mathcal #1}}

\begin{document}
\preprint{INT PUB-06-09}
\preprint{NT@UW-06-11}
\title{Dimer scattering in the $\epsilon$ expansion}

\author{Gautam Rupak
%\footnote{Email: {\tt grupak@u.washington.edu}} 
}
\email{grupak@u.washington.edu}
\affiliation{Institute for Nuclear Theory,
 University of Washington,
Seattle, WA 98195}
\affiliation{Nuclear Theory Group,
 University of Washington,
Seattle, WA 98195}

\begin{abstract}
The atom-dimer and dimer-dimer scattering lengths are 
analytically calculated in
an expansion around four spatial dimensions for fermions with a large
2-body scattering length $a$. We find the atom-dimer scattering length 
$a_{ad}/a = \frac{4}{3}-\frac{2}{9}\epsilon+\mcal O(\epsilon^2)$ and the
dimer-dimer scattering length $a_{dd}/a = \epsilon 
-0.344\epsilon^2+\mcal O(\epsilon^3)$, where $\epsilon= 4-d$ and $d$ is
the number of spatial dimensions. These ratios 
$a_{ad}/a = 1.11$ and $a_{dd}/a = 0.656$ 
at $\epsilon =1$ are to
be compared with the non-perturbatively calculated
numerical results $1.18$ and $0.6$
respectively. The neutron-deuteron scattering length in the quartet
channel $a_{nD}\approx 4.78$ fm is in reasonable agreement with the experimental
value $a^\mathrm{exp}_{nD}=6.35\pm0.02$ fm, considering 2-body effective range corrections 
$r_0/a\sim 0.4$ were excluded. 
The deuteron-deuteron scattering length 
$a_{DD}\approx3.15$ fm in the spin-$2$ channel. 
\end{abstract}

\maketitle

Systems with a large 2-body scattering length $a$ are common in
nuclear and atomic physics. Recent experiments near the Feshbach
resonance have added new systems to this 
list~\cite{JEThomas,TBourdel,KDieckmann,DJin,SJochim}. 
Near the unitarity limit
$a\rightarrow\pm \infty$, the leading behavior is expected to be
universal on dimensional ground. However, theoretical calculations in
this interesting regime of large 2-body scattering length are
difficult -- typically involving non-perturbative resummation of
diagrams or non-trivial numerical calculations. 
In Ref.~\cite{Nussinov}, Nussinov and Nussinov 
pointed out that in $d=4$ spatial
dimensions one obtains a non-interacting system of dimers. 
Recently Nishida and
Son have developed an $\epsilon= 4-d $ expansion around four spatial
dimensions~\cite{Nishida:2006br}, allowing them to perturbatively calculate
thermodynamic properties for a dilute Fermi gas at the unitarity limit
$a\rightarrow\pm\infty$. 

In this paper, we consider the calculation of the atom-dimer and
dimer-dimer scattering amplitude at large but finite 2-body $S$-wave scattering
length $a$ such that the
interaction is completely determined by 2-body physics. 
Besides providing a perturbative method for few-body calculations in
the 
vacuum, this work would allow study of many-body properties at finite
$a$ such as the Bose-Einstein condensation (BEC) in atomic systems near the 
Feshbach resonance. The relevant
theory for large scattering length $a$ is described by the following Lagrangian density:
\begin{align}\label{L0}
\mcal L =&\psi^\dagger\(i\partial_0+\frac{\nabla^2}{2M}\)\psi
-\frac{c_0}{4}\(\psi^T\sigma_2 \psi\)^\dagger(\psi^T\sigma_2 \psi)\ ,
\end{align}
where $\psi$ are spin-$\frac{1}{2}$ fermionic fields and the pauli
matrix $\sigma_2$ projects the interaction onto the singlet
$S$-wave. An equivalent form for the Lagrangian density after a
Hubbard-Stratonovich transformation is
\begin{align}\label{L1}
\mcal L=&\psi^\dagger\(i\partial_0+\frac{\nabla^2}{2M}\)\psi
+\frac{g}{2}\[\phi^\dagger\(\psi^T\sigma_2
\psi\)+h.c.\]\\
&+\frac{g^2}{c_0}\phi^\dagger\phi\ .\nonumber
\end{align}

\begin{figure}[tbh]
\includegraphics[width=0.47\textwidth,height=0.35\textwidth,clip]{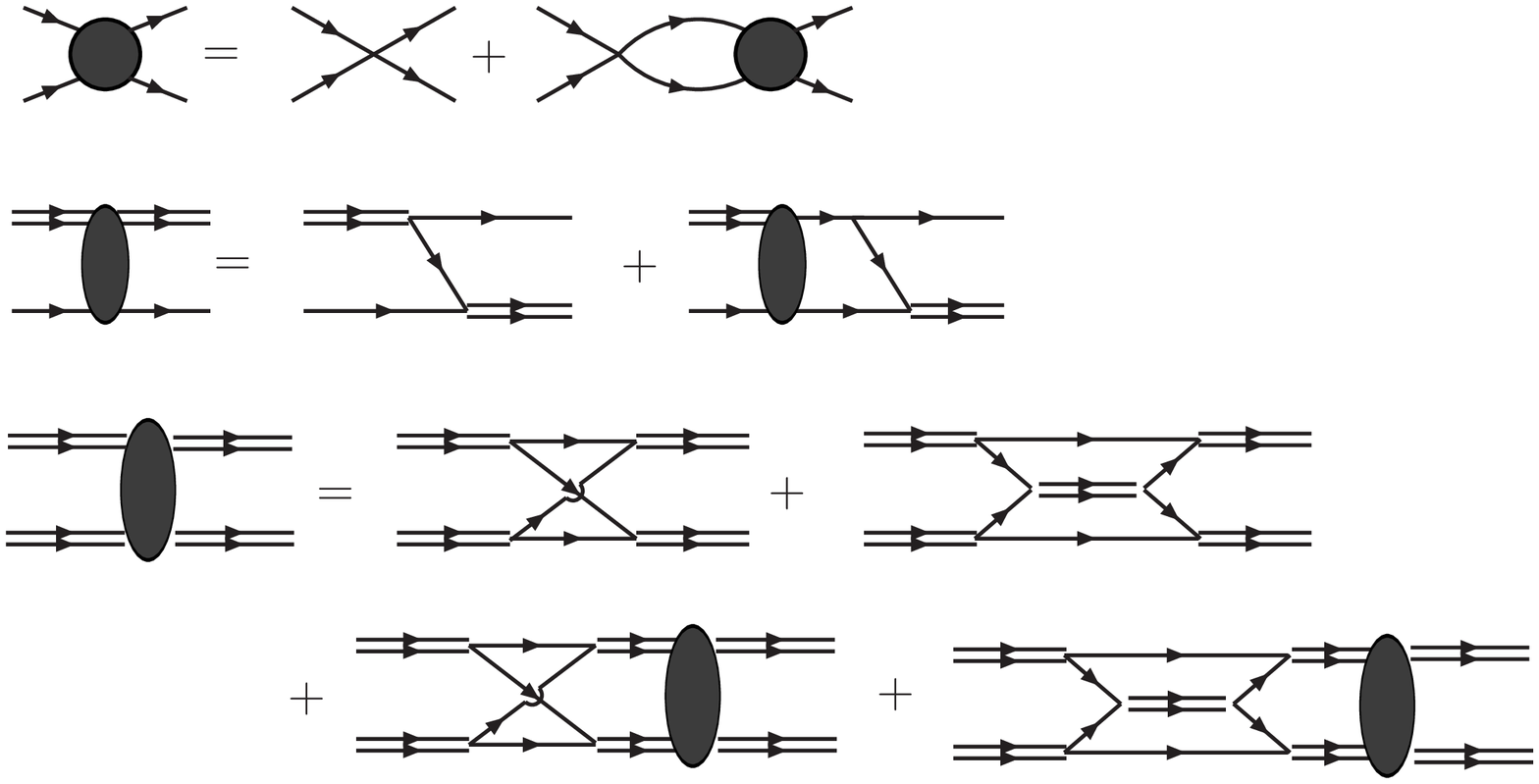}
\caption{\protect The \emph{leading order} 
scattering amplitudes for atoms and dimers in $d=3$ spatial 
dimensions. 
Single lines: atoms/fermions, double 
lines: \emph{fully dressed} dimers/bosons.}
\label{Scattering}
\end{figure}

In $d=3$ spatial dimensions, it is possible to establish a systematic
perturbation in $R/a$ where the range of the interaction 
$R\ll a$~\cite{Kaplan:1998tg,vanKolck:1998bw,Bedaque:1997qi,Chen:1999tn}. 
This has been quite successful in the 2 and 3-nucleon
systems, see for example 
Refs.~\cite{Bedaque:1998mb,Butler:2000zp,Rupak:1999rk}, 
 where the 2-body scattering length is large. However, it
involves resumming certain diagrams to all order in perturbation. 
In Fig.~\ref{Scattering}, the atom-atom scattering amplitude is $\mcal
O(a)$ and requires resumming an infinite set of 
diagrams all of which are leading order in the 
perturbation~\cite{Kaplan:1998tg,vanKolck:1998bw,Bedaque:1997qi,Chen:1999tn}. 
The
leading order 
atom-dimer amplitude is again $\mcal O(a)$ and given by an infinite set of
diagrams, see Ref.~\cite{Bedaque:1997qi}.  
The dimer-dimer amplitude is also $\mcal O(a)$, and similarly requires
resumming the indicated diagrams at leading order~\cite{LuuRupak}. 
The atom-dimer amplitude for spin-$\frac{1}{2}$ fermions has
been non-perturbatively calculated model-independently in effective field 
theory~\cite{Bedaque:1997qi,Bedaque:1998mb,Rupak:2006pu}. 
For dimer-dimer
scattering only a partial resummation of the diagrams has been done
in Ref.~\cite{Pieri:2000}. 

Working in $d=4$ spatial dimensions will avoid the
non-perturbative treatment of the dimer mentioned above~\cite{Nussinov}. 
 We will formulate an $\epsilon$
expansion for atom-dimer and dimer-dimer scattering in $d=4-\epsilon$
dimensions~\cite{Nishida:2006br}. First we define the coupling $c_0$ in arbitrary $d$
spatial dimensions using the pole in the 2-body scattering amplitude, 
Fig.~\ref{Scattering}
\begin{align}
i\mcal T_{aa}=\frac{-i c_0}{1+ic_0 L(p_0,\vec p)}
\end{align}
located at the 2-body binding energy $B = 1/(a^2 M)$:
\begin{align}
\frac{1}{c_0} =&-i L(p_0, \vec{p})\Big|_{p_0=-B+p^2/(4M)}\\
=&-\frac{M}{(4\pi)^{d/2}} \(M B\)^\frac{d-2}{2}
\Gamma\(1-\frac{d}{2}\),\nonumber
\end{align}
where 
\begin{align}
L(p_0,\vec p)=&-iM\int \frac{d^d\ q}{(2\pi)^d}\frac{1}{q^2-M p_0
  +\frac{p^2}{4}-i0^+}\\ 
=&-i\frac{M}{(4\pi)^{d/2}}\(-M
  p_0+\frac{p^2}{4}-i0^+\)^\frac{d-2}{2}\Gamma\(1-\frac{d}{2}\)
\nonumber, 
\end{align}
and then write the Lagrangian density in terms of the kinetic and
interaction piece as
\begin{align}\label{L2}
\mcal L=&\psi^\dagger\(i\partial_0+\frac{\nabla^2}{2M}\)\psi+
\phi^\dagger\(i\partial_0+\frac{\nabla^2}{4M}+B\)\phi +\mcal L_{I}\ , \\
\mcal L_I=&-\phi^\dagger\(i\partial_0+\frac{\nabla^2}{4M}+B\)\phi
 +\frac{g}{2}\[\phi^\dagger\(\psi^T\sigma_2
\psi\)+h.c.\] \nonumber\\
&+\frac{g^2}{c_0}\phi^\dagger\phi\ .\nonumber
\end{align}

\begin{figure}[tbh]
\includegraphics[width=0.45\textwidth,clip]{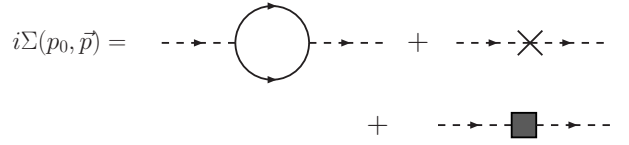}
\caption{\protect Dimer wavefunction renormalization. 
Solid lines: atoms/fermions, dashed lines: dimers/bosons. Unlike in
$d=3$ dimensions,  
the dimer propagators are not dressed since the dimer-fermion interaction is
perturbative around $d=4$. 
The
  ``X'' represents insertion of $ig^2/c_0$ and the ``square'' 
  kinetic energy $-i[p_0+B-\vec{p}^2/(2M)]$ from 
$\mcal L_I$ in Eq.~(\ref{L2}), on the dimer propagator.
}
\label{Zd}
\end{figure}

The original Lagrangian density in Eq.~(\ref{L0}) had only one coupling
$c_0$ that was determined by the 2-body scattering length $a$, the only
parameter in the theory. In Eq.~(\ref{L1}), after the
Hubbard-Stratonovich transformation, we have an additional
coupling $g$. However, its value is arbitrary and only the ratio of the
yukawa coupling squared $g^2$ and the ``mass'' term $g^2/c_0$, 
i.e. $c_0$, contributes after the dimer wavefunction renormalization. 
In $d=3$ dimensions, the wavefunction renormalization for the dimer
$Z_d\sim 1/a$~\cite{Chen:1999tn}. In $d=4-\epsilon$ dimensions, where we expect
a non-interacting theory of dimers~\cite{Nussinov}, the
coupling $g$ is chosen such that using the Lagrangian density in
Eq.~(\ref{L2}) gives $Z_d\sim 1 +\mcal O(\epsilon)$. This then requires
the dimer self-energy $i\Sigma(p_0, \vec p)$ in Fig.~\ref{Zd} to
vanish on-shell and only contribute to $Z_d$ through a derivative term at
$\mcal O(\epsilon)$:
\begin{align}\label{Zd_cond}
i\Sigma(p_0,\vec p)\approx&\epsilon\(p_0+B-\frac{p^2}{4M}\)
i\frac{\partial\Sigma}{\partial p_0}\Big|_{p_0=-B+p^2/(4M)} ,\\
Z_d=&1-\epsilon
\frac{\partial\Sigma}{\partial p_0}\Big|_{p_0=-B+p^2/(4M)}
+\cdots, \nonumber
\end{align}
where ``$\cdots$'' indicate terms higher order in the $\epsilon$
expansion. 

In $\epsilon=4-d$ dimensions, for Eq.~(\ref{Zd_cond}) to hold, 
 the coupling $g$ has to be 
\begin{align}
g=&\frac{\sqrt{8\pi^2\epsilon}}{M}
\underbrace{M^{\epsilon/2}}_\mathrm{arbitrary\ scale},
\end{align}
where only the factor $\frac{\sqrt{8\pi^2\epsilon}}{M}$ is uniquely 
determined from the dimer wavefunction renormalization. 
With this choice of the coupling $g$, it is straightforward
to estimate the relative sizes of diagrams. For elastic scattering,
diagrams involve even powers of $g$ and we associate a power of
$\epsilon$ for every factor of $g^2$. In effect, the $\epsilon$
expansion is a loop expansion.   

For the wavefunction renormalization $Z_d$, we get:
\begin{align}\label{EqZd}
i\Sigma(p_0, \vec p)=&-g^2 L(p_0,\vec{p})+ i\frac{g^2}{c_0}
-i\(p_0+B-\frac{p^2}{4M}\),\\
Z_d\approx&
1+\frac{\epsilon}{2}\[\gamma_E-\log(4\pi a^2 M^2)\], \nonumber
\end{align}
where $\gamma_E\approx 0.57722$ is the Euler-Mascheroni
constant.

\begin{figure}[tbh]
\includegraphics[width=0.45\textwidth,clip]{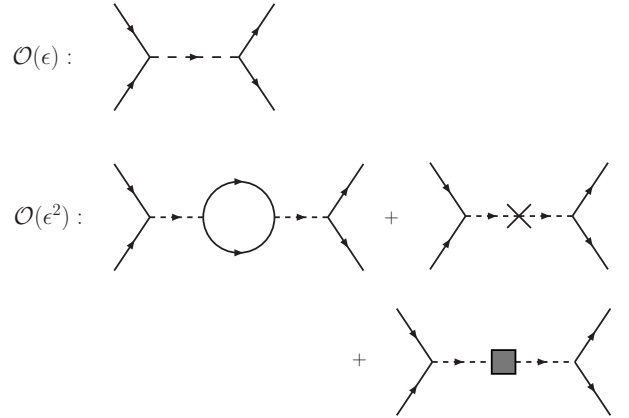}
\caption{\protect Atom-atom scattering amplitude $\mcal T_{aa}$ 
in the $\epsilon$
  expansion. Diagrams up to the next-to-leading order are shown. 
Same notation as in Fig.~\ref{Zd}.}
\label{AtomAtom}
\end{figure}

The atom-atom scattering amplitude is given by the diagrams in
Fig.~\ref{AtomAtom}. We find at threshold:
\begin{align}
 \frac{i\mathcal T_{aa}}{a^{3-\epsilon}}
 =&-i\frac{8\pi^2 }{a M}\epsilon
+i\frac{4\pi^2}{a M}\[1-\gamma_E +\log\(4\pi\)\]\epsilon^2+\cdots,
\end{align}
where we divided the amplitude $i\mcal T_{aa}$ by factors
$a^{3-\epsilon}$ to look at dimensionless ratios for convenience. 
Unlike in $d=3$ dimensions, the atom-atom scattering amplitude $i\mcal
T_{aa}$ 
scales as $\mcal O(a^2)$ in $d=4$ dimensions. 

\begin{figure}[tbh]
\includegraphics[width=0.45\textwidth,clip]{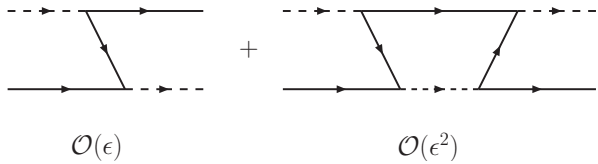}
\caption{\protect Atom-dimer scattering amplitude $\mcal T_{ad}$ in the $\epsilon$
  expansion. Diagrams up to the next-to-leading order are shown. Same
  notation as in Fig.~\ref{Zd}.}
\label{AtomDimer}
\end{figure}

A non-trivial check of the $\epsilon$ expansion with finite $a$ would
be the atom-dimer scattering length. 
Like the atom-atom scattering amplitude, the atom-dimer amplitude
$i\mcal T_{ad}$ 
also starts at $\mcal O(\epsilon)$, 
see Fig.~\ref{AtomDimer}. At $\mcal O(\epsilon^2)$, there is
a contribution from a $1$-loop diagram shown in Fig.~\ref{AtomDimer},
in addition to the wavefunction renormalization factor $Z_d$
calculated in Eq.~(\ref{EqZd}).    
The contribution from the tree diagram (including a factor of
$a^{\epsilon-3}$ ) in Fig.~\ref{AtomDimer} is:
\begin{align}
-i \frac{g^2 a^2 M Z_d}{ a^{3-\epsilon}}
=&-i\frac{8\pi^2}{a M}\epsilon 
-i\frac{4\pi^2 }{a M}\[\gamma_E-\log(4\pi)\]\epsilon^2+\cdots . 
\end{align}
Together with the contribution from the $1$-loop diagram, we get at
next-to-leading order
\begin{align}
\frac{i \mcal T_{ad}}{a^{3-\epsilon}}
=&-i\frac{8\pi^2}{a M}\epsilon
-i\frac{4\pi^2}{3 a M}\[3\gamma_E-4 -3\log(4\pi)\]\epsilon^2\ .
\end{align}
To compare with numerical results calculated in $3$ spatial
dimensions, we consider the dimensionless ratio
\begin{align}
\frac{\mcal T_{ad}}{\mcal T_{aa}}=& 1-\frac{\epsilon}{6}+\mcal O(\epsilon^2)
\approx0.83\Big|_{\epsilon =1}\ .
\end{align}
In $d=3$ spatial 
dimensions, $\mcal T_{ad}/\mcal T_{aa}=3 a_{ad}/(4 a)$. 
Numerical evaluation by solving the 3-body
Faddeev equation~\cite{Bedaque:1997qi,Bedaque:1998mb,Rupak:2006pu}
\begin{align}
\mcal T_{ad} =& \frac{3\pi a }{M} b(3 x^2/4-1, x,x)\Big|_{x=0}\ ,
\\
b(\eta, x, z)=& -K(\eta, x, z)
-\frac{2}{\pi}\int_0^\infty dy \frac{y^2 b(\eta,x,y)}
{y^2-\frac{4}{3}(\eta+1)} K(\eta, y, z)\ ,\nonumber\\
K(\eta,x,z)=&\frac{2}{3}\frac{1+\sqrt{\frac{3}{4}x^2-\eta}}{x
  z}
\log\[\frac{x^2+z^2+x z-\eta}{x^2+z^2-x z-\eta}\], \nonumber
\end{align}
gives $\mcal T_{ad}/\mcal T_{aa}\approx0.885$ ($a_{ad}\approx 1.18 a$).
The $\epsilon$ expansion result is perturbatively close to the 
non-perturbatively calculated numerical value, and it is a tremendous
simplification over the non-perturbative calculation in $d=3$
dimensions. 

The discussion can be generalized to certain spin channels in nuclear
physics where one has to introduce isospin degrees of freedom to
accommodate both the proton and the neutron with similar properties
under the strong nuclear interaction. Specifically, the dimer in nuclear
physics context is the deuteron, a bound state of a proton and a
neutron, 
 in the spin-1 channel. The spin degrees of freedom in the dimer
calculation can be thought of as the isospin of nuclear physics, and
then 
the dimer calculation can be applied to deuteron scattering in
channels where all the proton and neutron spins are pointing in the
same direction. Thus the atom-dimer amplitude is the same as
the neutron-deuteron scattering in the $S$-wave quartet channel with 
angular momentum $J=3/2$. 
 
The deuteron binding energy of $B=2.2246$ MeV
corresponds to $a=1/\sqrt{M
  B}=4.32$ fm, ignoring the effective range 
$r_0=1.75$ fm~\cite{deSwart:1995ui}. The
standard nuclear physics definition of the scattering length 
$\hat a \approx a+r_0/2\approx 5.2$ fm. From the atom-dimer 
calculation, the neutron-deuteron quartet channel scattering length
is  $a_{nD}\approx 1.11 a \approx 4.78$ fm. This is in reasonable agreement with  
the experimental value
$a^\mathrm{exp}_{nD}=6.35\pm0.02$ fm~\cite{Dilg} considering we neglected 
 $r_0/a\sim 0.4$ effects. For the deuteron, the effective range $r_0$
 corrections are significant 
 even
though formally higher order in the perturbation. The so-called
$Z$-parameterization was developed in $d=3$ dimensions to precisely
account for these effects~\cite{Phillips:1999hh}.     

\begin{figure}[tbh]
\includegraphics[width=0.45\textwidth,clip]{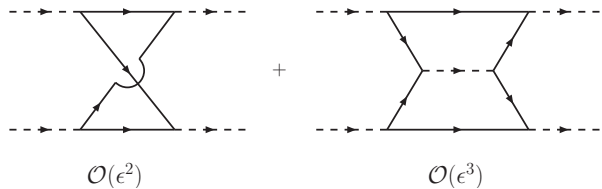}
\caption{\protect Dimer-dimer scattering amplitude $\mcal T_{dd}$ in the $\epsilon$
  expansion. Diagrams up to the next-to-leading order are shown. Same
  notation as in Fig.~\ref{Zd}.}
\label{DimerDimer}
\end{figure}

Finally, the dimer-dimer scattering length is calculated in the
$\epsilon$ expansion. 
The leading contribution to the dimer-dimer scattering is $\mcal
O(\epsilon^2)$ from the $1$-loop diagram shown in
Fig.~\ref{DimerDimer}. We calculate the dimer-dimer scattering
amplitude to next-to-leading order where it receives a $2$-loop
contribution. The $1$-loop contribution is: 
\begin{align}
-i &\frac{2 M^3 Z_d^2 g^4}{a^{3-\epsilon}}
\int\frac{d^d\ q}{(2\pi)^d}
\frac{1}{\[q^2+1/a^2\]^3}\\
&= -i\frac{2 M^3 Z_d^2 g^4}{(4\pi)^{d/2}}
\frac{a^{6-d}}{a^{3-\epsilon}}
\frac{\Gamma(3-d/2)}{\Gamma(3)}\nonumber\\
&=-i\frac{4\pi^2 }{a M}\epsilon^2
-i \frac{2\pi^2 }{a M}\[\gamma_E-\log(4\pi)\]\epsilon^3\nonumber 
+\cdots. 
\end{align}
The $2$-loop contribution gives
\begin{align}
i \frac{2 M^5 Z_d^2 g^6}{a^{3-\epsilon}}
&\int\frac{d^d\ q}{(2\pi)^d}\frac{d^d\ l}{(2\pi)^d}
\frac{1}{\[q^2+1/a^2\]^2\[l^2+1/a^2\]^2}\\
&{}\frac{1}{(q+l)^2/4+q^2/2+l^2/2+1/a^2}\nonumber\\
=&i\frac{16\pi^2 }{3 a M}\[1-18\log(3)+14\log(4)\]\epsilon^3 
+\mcal O(\epsilon^4)\ . \nonumber
\end{align}
Thus, for the next-to-leading order dimer-dimer amplitude, we get the
analytic result:
\begin{align}
\frac{i\mcal T_{dd}}{a^{3-\epsilon}}=& -i\frac{4\pi^2 }{a M}\epsilon^2
+i\frac{2\pi^2 }{3 a M}\[8-3\gamma_E
+224\log(2)\right.\\
&\left.-144\log(3)+3\log(4\pi) \]\epsilon^3\ ,\nonumber\\
\frac{\mcal T_{dd}}{\mcal T_{aa}}=&\frac{1}{2}\epsilon
-0.172\epsilon^2+\mcal O(\epsilon^3)
\approx 0.328\Big|_{\epsilon=1}\ . \nonumber
\end{align}
$\mcal T_{dd}/\mcal T_{aa}= a_{dd}/{2a}$ in $d=3$ spatial
dimensions and numerical calculation gives $a_{dd}=0.6 a$~\cite{Petrov}, a
value perturbatively reproduced in the epsilon expansion. From the
dimer-dimer calculation, the deuteron-deuteron scattering length in 
channels with all the proton and neutron spins in the same direction is 
$a_{DD}\approx 3.15$ fm in the $\epsilon$ expansion, ignoring 2-body
effective range $r_0$ effects. 

Systems with unnaturally large 2-body scattering lengths are common in
nature and typically necessitate non-perturbative calculations that
are computationally difficult and expensive. 
Study and development of perturbative 
techniques are always useful. Many times perturbative calculations can be
done analytically that allow insights otherwise not possible in a
numerical calculation.   
In this paper, we considered a fermi system with a large
but finite 
2-body scattering length $a$ in an expansion around $d=4$ spatial
dimensions. The effect of the 2-body scattering length was
systematically incorporated and the atom-dimer and dimer-dimer
scattering length was calculated to next-to-leading order in
perturbation. The results seem to converge to the  non-perturbatively 
calculated numerical 
solutions at this order of the calculation, and the $\epsilon$
expansion should be an useful tool for studying strongly interacting
systems with large 2-body scattering length $a$. For example, the current work
would be important for research in the BEC-BCS
crossover region in
atomic systems as one varies the scattering length $a$ across the
Feshbach resonance. As shown, the dimer calculation can be generalized
to deuterons in nuclear physics in certain spin channels. 
However, it is not clear from this calculation how the Efimov 
effect~\cite{Efimov} in
bosons and fermions with isospin (more than one fermion species as in
the 
proton-neutron system) associated with non-perturbative
renormalization~\cite{Bedaque:1998kg} emerges in the $\epsilon$ expansion.  
It could be that the Efimov effect cannot be described by an analytic
continuation from $d=4$ spatial dimensions. It is known that the Efimov effect
occur only in spatial 
dimensions $2.3 < d < 3.8$, with $d=3$ being
the only integer dimension~\cite{Nielson}. 

\begin{acknowledgments}
The author thanks D. T. Son for helpful discussions, and P. F. Bedaque
and D. B. Kaplan  for comments on the manuscript. 
This work was supported in part by DOE grants DE-FG02-00ER41132, 
DE-FC02-01ER41187 and DE-FG02-97ER41014.
\end{acknowledgments}
%==============Bibliography =====================
%\bibliographystyle{h-physrev4}
%\bibliography{AtomDimer}

\end{document}